\documentclass[nofootinbib,superscriptaddress,notitlepage,pre]{revtex4-1}%
\usepackage{amssymb}
\usepackage{amsfonts}
\usepackage{amsmath}
\usepackage{graphicx}
\usepackage{color}
\usepackage{version}
\usepackage{numinsec}%
\providecommand{\U}[1]{\protect\rule{.1in}{.1in}}

\excludeversion{Commentskip}
\excludeversion{Commentdel}
\excludeversion{Commentnew}

\begin{document}
\title[Phase space dynamics]{Wigner distribution functions for complex dynamical systems: a path integral approach}
\author{Dries Sels}
\email{Corresponding author: dries.sels@ua.ac.be}
\affiliation{Physics Department, University of Antwerp, Universiteitsplein 1, 2060
Antwerpen, Belgium}
\author{Fons Brosens}
\email{fons.brosens@ua.ac.be}
\affiliation{Physics Department, University of Antwerp, Universiteitsplein 1, 2060
Antwerpen, Belgium}
\author{Wim Magnus }
\email{wim.magnus@ua.ac.be}
\affiliation{Physics Department, University of Antwerp, Universiteitsplein 1, 2060
Antwerpen, Belgium}
\affiliation{MSP, imec, Kapeldreef 75, 3001 Leuven, Belgium}

\begin{abstract}
Starting from Feynman's Lagrangian description of quantum mechanics, we
propose a method to construct explicitly the propagator for the Wigner
distribution function of a single system. For general quadratic Lagrangians,
only the classical phase space trajectory is found to contribute to the
propagator. Inspired by Feynman's and Vernon's influence functional theory we
extend the method to calculate the propagator for the reduced Wigner function
of a system of interest coupled to an external system. Explicit expressions
are obtained when the external system consist{s} of a set of independent
harmonic oscillators. As an example we calculate the propagator for the
reduced Wigner function associated with the Caldeira-Legett model.

\end{abstract}
\volumeyear{year}
\volumenumber{number}
\issuenumber{number}
\eid{identifier}
\date{\today}
\maketitle

\section{Introduction}

To explain the dynamic behavior of a quantum mechanical system of interest one
generally needs to extract the time dependence of the density matrix. Then, an
arbitrary observable represented by a Hermitian operator can be obtained by
taking the trace of the operator times the density matrix.

In a classical description, however, the density matrix emerges as a
continuous, time dependent phase space distribution function whereas physical
observables are identified with ordinary functions. Multiplying the latter
with the phase space distribution function and integrating over phase space
yields the commonly known recipe to calculate all relevant expectation values.
Both approaches can be linked if one addresses the Wigner
function\cite{Wigner} which is nothing but the Wigner--Weyl transform of the
quantum mechanical density matrix. Inheriting from the density matrix all
necessary information to calculate any observables of interest, the Wigner
satisfies the Wigner-Liouville equation, the quantum mechanical analog of the
classical Liouville equation.

Alternative to the Hamiltonian description of quantum mechanics one may adopt
Feynman's Lagrangian formulation of quantum mechanics\cite{FeynmSpacetime}.
This approach essentially relies on the the calculation of the Green's
function or propagator appearing as a path integral, rather than on solving
the dynamical equations equation for the density matrix. Moreover, also
Feynman's approach exhibits a direct link to classical mechanics through the
explicit occurrence of the classical action in the path integral.

In this work, adopting exclusively Feynman's Lagrangian formulation, we
construct the path integral for the propagator of the single particle Wigner
function in section \ref{sec:Propagator}. In section
\ref{sec:Influencefunctional} we extend the obtained result by computing also
the path integral for the propagator of the reduced Wigner function of a
system coupled to an external quantum system. This extension may be seen as
the Wigner-Weyl formulation of the theory of influence functionals developed
by Feynman and Vernon\cite{FeyVer}. In section \ref{sec:Harmonicinfluence} we
derive the explicit Wigner influence functional for a particle coupled to a
set of independent harmonic oscillators. Finally, as a non trivial example, we
calculate the Wigner function propagator of a particle described by the
Caldeira-Legett model\cite{CaldeiraLeggett}.

\section{Single-particle phase space propagator\label{sec:Propagator}}

The well-known Wigner distribution function~\cite{Wigner} provides a phase
space description of quantum mechanics, and its dynamics is usually formulated
as an inhomogeneous partial integro-differential equation. In earlier
work~\cite{BMNewton,SBMFeff,SBMpathintegral} we investigated the relation
between this equation of motion and classical phase space trajectories.
However, inspired by Feynman's Lagrangian description of quantum
mechanics~\cite{FeynmSpacetime}, in this paper we have explicitly constructed
a propagator (i.e., a Green's function) for the phase space distribution.

Consider first a single-particle system, for simplicity in the notations in
one dimension,whose coordinates are denoted by $x$\textbf{. }The
quantum-mechanical amplitude for the system to go from position $x_{a}$ at
time $t=t_{a}$ to position $x_{b}$ at time $t=t_{b}$ is given by the Feynman
path integral%
\begin{equation}
K\left(  x_{b},t_{b}|x_{a},t_{a}\right)  =\int_{x(t_{a})=x_{a}}^{x(t_{b}%
)=x_{b}}\mathcal{D}x\left(  t\right)  \exp\left(  \frac{i}{\hbar}S\left[
x\left(  t\right)  \right]  \right)  , \label{eq:FeynmanPI1}%
\end{equation}
where $S\left[  x\left(  t\right)  \right]  $ is the action of the system for
a trajectory $x\left(  t\right)  $. Let $\left\{  \left\vert n\right\rangle
\right\}  $ denote a complete orthonormal set of states. The amplitude
$A\left(  n_{b},t_{b}|n_{a},t_{a}\right)  $ for being in state $\left\vert
n_{b}\right\rangle $ at time $t_{b},$ if initially in state $\left\vert
n_{a}\right\rangle $ at time $t_{a},$ is then given by
\begin{equation}
A\left(  n_{b},t_{b}|n_{a},t_{a}\right)  =%
{\displaystyle\iint}
\left\langle n_{b}|x_{b}\right\rangle K\left(  x_{b},t_{b}|x_{a},t_{a}\right)
\left\langle x_{a}|n_{a}\right\rangle \text{\textrm{d}}x_{a}\text{d}x_{b}.
\label{eq:Amplitude_nbna}%
\end{equation}
The corresponding transition probability $P\left(  n_{b},t_{b}|n_{a}%
,t_{a}\right)  =\left\vert A\left(  n_{b},t_{b}|n_{a},t_{a}\right)
\right\vert ^{2}$ can thus be written as%
\begin{equation}
P\left(  n_{b},t_{b}|n_{a},t_{a}\right)  =%
{\displaystyle\iiiint}
\left\langle x_{b}^{\prime}|n_{b}\right\rangle \left\langle n_{b}%
|x_{b}\right\rangle K^{\ast}\left(  x_{b}^{\prime},t_{b}|x_{a}^{\prime}%
,t_{a}\right)  K\left(  x_{b},t_{b}|x_{a},t_{a}\right)  \left\langle
x_{a}|n_{a}\right\rangle \left\langle n_{a}|x_{a}^{\prime}\right\rangle
\mathrm{d}x_{a}\mathrm{d}x_{b}\mathrm{d}x_{a}^{\prime}\mathrm{d}x_{b}^{\prime
}. \label{eq:Probability_nbna}%
\end{equation}
Therefore the total probability to be in state $\left\vert n_{b}\right\rangle
$ at $t=t_{b}$ is the sum over all possible transitions from $\left\vert
n_{a}\right\rangle $ to $\left\vert n_{b}\right\rangle ,$ weighted by the
initial probability to be in a given state $\left\vert n_{a}\right\rangle $:%
\begin{equation}
P(n_{b},t_{b})=\sum_{n_{a}}P(n_{a},t_{a})P\left(  n_{b},t_{b}|n_{a}%
,t_{a}\right)  ,
\end{equation}
such that%
\begin{equation}
P(n_{b},t_{b})=%
{\displaystyle\iiiint}
\left\langle x_{b}^{\prime}|n_{b}\right\rangle \left\langle n_{b}%
|x_{b}\right\rangle K^{\ast}\left(  x_{b}^{\prime},t_{b}|x_{a}^{\prime}%
,t_{a}\right)  K\left(  x_{b},t_{b}|x_{a},t_{a}\right)  \left\{  \sum_{n_{a}%
}P(n_{a},t_{a})\left\langle x_{a}|n_{a}\right\rangle \left\langle n_{a}%
|x_{a}^{\prime}\right\rangle \right\}  \mathrm{d}x_{a}\mathrm{d}%
x_{b}\mathrm{d}x_{a}^{\prime}\mathrm{d}x_{b}^{\prime},
\label{eq:ProbabilityM1}%
\end{equation}
where the term between braces is identified as the initial density matrix
$\rho\left(  x_{a},x_{a}^{\prime},t_{a}\right)  $ of the system, since%
\begin{equation}
\rho\left(  x,x^{\prime},t\right)  =\sum_{n}\left\langle x|n\right\rangle
P(n,t)\left\langle n|x^{\prime}\right\rangle .
\end{equation}
Because of the orthonormality of the states, the probability to be in a state
$\left\vert n_{b}\right\rangle $ at time $t_{b}$ can also be extracted from
the density matrix by%
\begin{equation}
P(n_{b},t_{b})=%
{\displaystyle\iint}
\left\langle x_{b}^{\prime}|n_{b}\right\rangle \left\langle n_{b}%
|x_{b}\right\rangle \rho\left(  x_{b},x_{b}^{\prime},t_{b}\right)
\mathrm{d}x_{b}\mathrm{d}x_{b}^{\prime}. \label{eq:Probability_nb}%
\end{equation}
It follows, by comparison with Eq.(\ref{eq:ProbabilityM1}), that the density
matrix at time $t_{b}$ is given by%
\begin{equation}
\rho\left(  x_{b},x_{b}^{\prime},t_{b}\right)  =%
{\displaystyle\iint}
\text{ }K\left(  x_{b},t_{b}|x_{a},t_{a}\right)  K^{\ast}\left(  x_{b}%
^{\prime},t_{b}|x_{a}^{\prime},t_{a}\right)  \text{ }\rho\left(  x_{a}%
,x_{a}^{\prime},t_{a}\right)  \text{\textrm{d}}x_{a}\text{\textrm{d}}%
x_{a}^{\prime}. \label{eq:densitymatrixpropdef}%
\end{equation}
The density matrices can be expressed in terms of the Wigner distribution
function by an inverse Weyl transform%
\begin{equation}
f\left(  x,p,t\right)  =\int\rho\left(  x+\frac{\xi}{2},x-\frac{\xi}%
{2},t\right)  e^{-ip\xi/\hbar}\frac{\mathrm{d}\xi}{2\pi\hbar}~\leftrightarrow
~\rho\left(  x,x^{\prime},t\right)  =\int f\left(  \frac{x+x^{\prime}}%
{2},p,t\right)  e^{+ip\left(  x-x^{\prime}\right)  /\hbar}\mathrm{d}p.
\label{eq:Wignerdefinition}%
\end{equation}
Defining the propagator $K_{w}$ of the Wigner function as%
\begin{equation}
f\left(  x_{b},p_{b},t_{b}\right)  =%
{\displaystyle\iint}
K_{w}\left(  x_{b},p_{b},t_{b}|x_{a},p_{a},t_{a}\right)  \text{ }f\left(
x_{a},p_{a},t_{a}\right)  \text{\textrm{d}}x_{a}\text{\textrm{d}}p_{a},
\label{eq:Kw_def}%
\end{equation}
one may extract it from Eq.~(\ref{eq:densitymatrixpropdef}):%

\begin{align}
K_{w}\left(  x_{b},p_{b},t_{b}|x_{a},p_{a},t_{a}\right)   &  =%
{\displaystyle\iint}
e^{-\frac{i}{\hbar}\left(  p_{b}\xi_{b}-p_{a}\xi_{a}\right)  }K\left(
x_{b}+\frac{\xi_{b}}{2},t_{b}|x_{a}+\frac{\xi_{a}}{2},t_{a}\right)  K^{\ast
}\left(  x_{b}-\frac{\xi_{b}}{2},t_{b}|x_{a}-\frac{\xi_{a}}{2},t_{a}\right)
\frac{\text{\textrm{d}}\xi_{b}\text{\textrm{d}}\xi_{a}}{2\pi\hbar
}\label{eq:PropWignerGeneral_0}\\
&  =%
{\displaystyle\iint}
\int_{x(t_{a})=x_{a}}^{x(t_{b})=x_{b}}\int_{\xi(t_{a})=\xi_{a}}^{\xi
(t_{b})=\xi_{b}}\mathcal{D}x\mathcal{D}\xi\exp\left(  \frac{i}{\hbar}\left(
-p\xi|_{t_{a}}^{t_{b}}+S\left[  x\mathbf{+}\frac{\xi}{2}\right]  -S\left[
x-\frac{\xi}{2}\right]  \right)  \right)  \frac{\text{\textrm{d}}\xi
_{b}\text{\textrm{d}}\xi_{a}}{2\pi\hbar}. \label{eq:PropWignerGeneral}%
\end{align}
Note that the explicit time dependence in the path variables $x\left(
t\right)  $ and $\xi\left(  t\right)  $ is omitted in order not to overload
the notations.

For example, consider $x$ to be the coordinate of a (nonrelativistic) particle
with mass $m$ in an external potential $V(x),$ described by the action
\begin{equation}
S\left[  x\right]  =\int_{t_{a}}^{t_{b}}\left(  \frac{m}{2}\dot{x}%
^{2}-V(x)\right)  \mathrm{d}t.
\end{equation}
One then finds%
\[
S\left[  x\mathbf{+}\frac{\xi}{2}\right]  -S\left[  x-\frac{\xi}{2}\right]
=\int_{t_{a}}^{t_{b}}\left(  m\dot{x}\dot{\xi}\mathbf{-}V\left(
x\mathbf{+}\frac{\xi}{2}\right)  +V\left(  x\mathbf{-}\frac{\xi}{2}\right)
\right)  \text{{\normalsize d}}t.
\]
After an integration by parts for the kinetic term one obtains the following
expression for the propagator $K_{w}:$%
\begin{multline*}
K_{w}\left(  x_{b},p_{b},t_{b}|x_{a},p_{a},t_{a}\right)  =%
{\displaystyle\iint}
\int_{x(t_{a})=x_{a}}^{x(t_{b})=x_{b}}\int_{\xi(t_{a})=\xi_{a}}^{\xi
(t_{b})=\xi_{b}}\mathcal{D}x\mathcal{D}\xi\exp\left(  \frac{i}{\hbar}\left(
m\dot{x}-p\right)  \xi|_{t_{a}}^{t_{b}}\right) \\
\times\exp\left(  -\frac{i}{\hbar}\int_{t_{a}}^{t_{b}}\left(  m\ddot{x}%
\xi\mathbf{+}V\left(  x\mathbf{+}\frac{\xi}{2}\right)  -V\left(
x\mathbf{-}\frac{\xi}{2}\right)  \right)  \text{{\normalsize d}}t\right)
\frac{\text{\textrm{d}}\xi_{b}\text{\textrm{d}}\xi_{a}}{2\pi\hbar}.
\end{multline*}
The path integral over $\xi$ sums over all possible paths between two fixed
points $\left(  \xi_{a},\xi_{b}\right)  ,$ but additionally one has to
integrate over all possible initial and final points $\left(  \xi_{a},\xi
_{b}\right)  .$ Performing this last integral results in an unconstrained path
integral over all possible paths. Moreover, the integral over the boundary
term fixes the initial and final momentum $p$ to $m\dot{x}\mathbf{,}$ such
that one is left with
\begin{equation}
K_{w}\left(  x_{b},p_{b},t_{b}|x_{a},p_{a},t_{a}\right)  =\frac{1}{2\pi\hbar
}\int_{\substack{x(t_{a})=x_{a}\\p(t_{a})=m\dot{x}_{a}}}^{\substack{x(t_{b}%
)=x_{b}\\p(t_{b})=m\dot{x}_{b}}}\mathcal{D}x\int\mathcal{D}\xi\exp\left(
-\frac{i}{\hbar}\int_{t_{a}}^{t_{b}}\left(  m\ddot{x}\xi\mathbf{+}V\left(
x\mathbf{+}\frac{\xi}{2}\right)  -V\left(  x\mathbf{-}\frac{\xi}{2}\right)
\right)  \text{\textrm{d}}t\right)  . \label{eq:PropWignerPotential}%
\end{equation}
Note that in previous work we obtained the same expession (Eq.~(7) in~
\cite{SBMpathintegral}) for the propagator of the Wigner function by an
infinitesimal time lapse expansion of the Wigner-Liouville equation. With this
alternative derivation we have shown that the Wigner function can also be
constructed directly from the Lagrangian form of quantum mechanics without the
need of the usual canonical quantization procedure. This is important because
some problems are much more complicated in Hamiltonian form, or a Hamiltonian
description might not even exist. Furthermore, we have shown that the
propagator~(\ref{eq:PropWignerPotential}) for the Wigner function is
completely general, and not restricted to the phase space evolution of pure states.

It is clear that the double path integral~(\ref{eq:PropWignerPotential}) can
only be solved analytically for a limited set of problems. As an example of
such a problem consider a particle in a time dependent quadratic potential
$V(x,t)=a(t)+b(t)x+c(t)x^{2}.$ In that case the propagator can be explicitly
evaluated:%
\begin{multline}
\underset{\text{quad}}{K_{w}}\left(  x_{b},p_{b},t_{b}|x_{a},p_{a}%
,t_{a}\right)  =\int_{\substack{x(t_{a})=x_{a}\\p(t_{a})=m\dot{x}_{a}%
}}^{\substack{x(t_{b})=x_{b}\\p(t_{b})=m\dot{x}_{b}}}\mathcal{D}%
x\int\mathcal{D}\xi\frac{\exp\left(  -\frac{i}{\hbar}\int_{t_{a}}^{t_{b}%
}\left(  m\ddot{x}\mathbf{+}b\left(  t\right)  +2c\left(  t\right)  x\right)
\xi\text{\textrm{d}}t\right)  }{2\pi\hbar}\label{eq:Harmonicexample}\\
\propto\int_{\substack{q(t_{a})=x_{a}\\p(t_{a})=m\dot{x}_{a}}%
}^{\substack{x(t_{b})=x_{b}\\p(t_{b})=m\dot{x}_{b}}}\mathcal{D}x\text{ }%
\delta\left(  m\ddot{x}\mathbf{+}b\left(  t\right)  +2c\left(  t\right)
x\right) \\
=\delta\left(  x_{b}-x_{\text{cl}}\left(  t_{b}|x_{a},p_{a},t_{a}\right)
\right)  \delta\left(  p_{b}-p_{\text{cl}}\left(  t_{b}|x_{a},p_{a}%
,t_{a}\right)  \right)  ),
\end{multline}
where $x_{\text{cl}}\left(  t_{b}|x_{a},p_{a},t_{a}\right)  $ and
$p_{\text{cl}}\left(  t_{b}|x_{a},p_{a},t_{a}\right)  $ are the position and
momentum at time $t_{b}$ along the classical trajectory with initial position
$x_{a}$ and momentum $p_{a}$ at time $t_{a}.$ This confirms the well known
result~\cite{BMNewton,Schleich} that the Wigner-Liouville equation can exactly
be solved by the method of characteristics for all harmonic problems. It is
the basic formula for most of the subsequent calculations. In the appendix a
detailed derivation is given, based on the Feynman propagator for this type of potentials.

\section{Influence functionals in phase space\label{sec:Influencefunctional}}

In the previous section we derived the phase space propagator for a
single-particle system. But many interesting systems consist of interacting
subsystems. One of those subsystems is usually of particular interest. In this
section we generalize the previous phase space description with the help of
influence functionals~\cite{FeyVer}, allowing to describe the behavior of the
subsystem of interest, coupled with an external (quantum) system, solely in
terms of its own variables.

Consider a more complicated system that consists of two subsystems with
coordinates $x$ and $u\mathbf{.}$ The subsystems are coupled by a potential
$V_{I}(x\mathbf{,}u),$ which is incorporated in the action as $S_{I}\left[
x,u\right]  .$

The extension of the path integral~(\ref{eq:FeynmanPI1}) thus takes the form%
\begin{equation}
\text{ }K\left(  x_{b},u_{b},t_{b}|x_{a},u_{a},t_{a}\right)  =\int
_{x(t_{a})=x_{a}}^{x(t_{b})=x_{b}}\int_{u(t_{a})=u_{a}}^{u(t_{b})=u_{b}%
}\mathcal{D}x\mathcal{D}u\exp\left(  \frac{i}{\hbar}\left(  S_{x}\left[
x\right]  +S_{u}\left[  u\right]  +S_{I}\left[  x,u\right]  \right)  \right)
\label{eq:Totalproprx}%
\end{equation}
where $S_{x}\left[  x\right]  $ and $S_{u}\left[  u\right]  $ are the
non-interacting contributions of the subsystems $x$ and $u\ $to the action.
Again, for brevity, the path variables $x$ and $u$ are implicitly assumed to
be time dependent. If $\left\{  \left\vert n\right\rangle \right\}  $ denotes
a complete orthonormal set of states for the $x$ subsystem, and $\left\{
\left\vert j\right\rangle \right\}  $ similarly for the $u$ subsystem, the
amplitude~(\ref{eq:Amplitude_nbna}) can be generalized to%
\begin{equation}
A\left(  n_{b},j_{b},t_{b}|n_{a},j_{a},t_{a}\right)  =%
{\displaystyle\iiiint}
\left\langle n_{b}|x_{b}\right\rangle \left\langle j_{b}|u_{b}\right\rangle
K\left(  x_{b},u_{b},t_{b}|x_{a},u_{a},t_{a}\right)  \left\langle u_{a}%
|j_{a}\right\rangle \left\langle x_{a}|n_{a}\right\rangle \text{\textrm{d}%
}x_{a}\text{d}x_{b}\text{\textrm{d}}u_{a}\text{d}u_{b},
\end{equation}
The conditional transition probability $P\left(  n_{b},j_{b},t_{b}|n_{a}%
,j_{a},t_{a}\right)  =$ $\left\vert A\left(  n_{b},j_{b},t_{b}|n_{a}%
,j_{a},t_{a}\right)  \right\vert ^{2}$ for subsystem $x$ to go from state
$\left\vert n_{a}\right\rangle $ at $t_{a}$ to state $\left\vert
n_{b}\right\rangle $ at $t_{b},$ while subsystem $u$ goes from $\left\vert
j_{a}\right\rangle $ to $\left\vert j_{b}\right\rangle $ thus becomes%
\begin{multline}
P\left(  n_{b},j_{b},t_{b}|n_{a},j_{a},t_{a}\right)  =%
{\displaystyle\iiiint}
{\displaystyle\iiiint}
\left\langle x_{b}^{\prime}|n_{b}\right\rangle \left\langle u_{b}^{\prime
}|j_{b}\right\rangle K^{\ast}\left(  x_{b}^{\prime},u_{b}^{\prime},t_{b}%
|x_{a}^{\prime},u_{a}^{\prime},t_{a}\right)  \left\langle j_{a}|u_{a}^{\prime
}\right\rangle \left\langle n_{a}|x_{a}^{\prime}\right\rangle \times
\label{eq:Pconditional}\\
\times\left\langle n_{b}|x_{b}\right\rangle \left\langle j_{b}|u_{b}%
\right\rangle K\left(  x_{b},u_{b},t_{b}|x_{a},u_{a},t_{a}\right)
\left\langle u_{a}|j_{a}\right\rangle \left\langle x_{a}|n_{a}\right\rangle
\text{\textrm{d}}x_{a}\text{d}x_{b}\text{\textrm{d}}u_{a}\text{d}%
u_{b}\text{\textrm{d}}x_{a}^{\prime}\text{d}x_{b}^{\prime}\text{\textrm{d}%
}u_{a}^{\prime}\text{d}u_{b}^{\prime}.
\end{multline}
From hereof we assume that only the description of subsytem $x$ is physically
relevant. In other words, one is interested in the probability $P\left(
n_{b},t_{b}|n_{a},t_{a}\right)  $ of the transition of subsystem $x$ from
state $\left\vert n_{a}\right\rangle $ to $\left\vert n_{b}\right\rangle .$
This can be found from the conditional probability~(\ref{eq:Pconditional}) by
summing over all final states $\left\vert j_{b}\right\rangle $ and initial
states $\left\vert j_{a}\right\rangle ,$ weighted by the probability that
subsystem $u$ was initially in state $\left\vert j_{a}\right\rangle :$%
\begin{equation}
P\left(  n_{b},t_{b}|n_{a},t_{a}\right)  =\sum_{j_{a},j_{b}}P(j_{a}%
,t_{a})P\left(  n_{b},j_{b},t_{b}|n_{a},j_{a},t_{a}\right)  .
\label{eq:Probability_nbna_from_nbjbnaja}%
\end{equation}
Substitution of the conditional probability~(\ref{eq:Pconditional}) into
Eq.~(\ref{eq:Probability_nbna_from_nbjbnaja}) and regrouping terms leads to%
\begin{multline}
P\left(  n_{b},t_{b}|n_{a},t_{a}\right)  =%
{\displaystyle\iiiint}
{\displaystyle\iiiint}
\left\langle x_{b}^{\prime}|n_{b}\right\rangle K^{\ast}\left(  x_{b}^{\prime
},u_{b}^{\prime},t_{b}|x_{a}^{\prime},u_{a}^{\prime},t_{a}\right)
\left\langle n_{b}|x_{b}\right\rangle \left\langle n_{a}|x_{a}^{\prime
}\right\rangle K\left(  x_{b},u_{b},t_{b}|x_{a},u_{a},t_{a}\right)
\left\langle x_{a}|n_{a}\right\rangle \\
\times\left\{  \sum_{j_{a}}\left\langle u_{a}|j_{a}\right\rangle P(j_{a}%
,t_{a})\left\langle j_{a}|u_{a}^{\prime}\right\rangle \right\}  \left\{
\sum_{j_{b}}\left\langle u_{b}^{\prime}|j_{b}\right\rangle \left\langle
j_{b}|u_{b}\right\rangle \right\}  \text{\textrm{d}}x_{a}\text{d}%
x_{b}\text{\textrm{d}}u_{a}\text{d}u_{b}\text{\textrm{d}}x_{a}^{\prime
}\text{d}x_{b}^{\prime}\text{\textrm{d}}u_{a}^{\prime}\text{d}u_{b}^{\prime}.
\end{multline}
The sum over the initial states $\left\vert j_{a}\right\rangle $ is clearly
identified as the initial density matrix of subsystem \textbf{$u$}$\mathbf{,}$
whereas the closure relation ensures that the sum over all final states
$\left\vert j_{b}\right\rangle $ reduces to $\delta(u_{b}-u_{b}^{\prime}).$
Then, substitution of expression~(\ref{eq:Totalproprx}) for the propagators
$K\left(  \cdots\right)  $ and rearranging terms one ends up with the
following expression for the required transition probability:%
\begin{multline}
P\left(  n_{b},t_{b}|n_{a},t_{a}\right)  =%
{\displaystyle\iiiint}
\left\langle x_{b}^{\prime}|n_{b}\right\rangle \left\langle n_{b}%
|x_{b}\right\rangle \left\langle n_{a}|x_{a}^{\prime}\right\rangle
\left\langle x_{a}|n_{a}\right\rangle \times\\
\times\int_{x(t_{a})=x_{a}}^{x(t_{b})=x_{b}}\int_{x^{\prime}(t_{a}%
)=x_{a}^{\prime}}^{x^{\prime}(t_{b})=x_{b}^{\prime}}\mathcal{D}x\mathcal{D}%
x^{\prime}\exp\left(  \frac{i}{\hbar}\left(  S_{x}\left[  x\right]
-S_{x}\left[  x^{\prime}\right]  \right)  \right)  \mathcal{F}\left[
x,x^{\prime}\right]  \text{\textrm{d}}x_{a}\text{d}x_{b}\text{\textrm{d}}%
x_{a}^{\prime}\text{d}x_{b}^{\prime},
\end{multline}
with%
\begin{multline}
\mathcal{F}\left[  x,x^{\prime}\right]  \mathcal{=}\iiiint\rho\left(
u_{a},u_{a}^{\prime},t_{a}\right)  \delta(u_{b}-u_{b}^{\prime})\left\{
\int_{u(t_{a})=u_{a}}^{u(t_{b})=u_{b}}\mathcal{D}u\exp\left(  \frac{i}{\hbar
}\left(  S_{u}\left[  u\right]  +S_{I}\left[  x,u\right]  \right)  \right)
\right\}  \times\label{eq:influencefunc}\\
\times\left\{  \int_{u^{\prime}(t_{a})=u_{a}^{\prime}}^{u^{\prime}%
(t_{b})=u_{b}^{\prime}}\mathcal{D}u^{\prime}\exp\left(  \frac{-i}{\hbar
}\left(  S_{u}\left[  u^{\prime}\right]  +S_{I}\left[  x^{\prime},u^{\prime
}\right]  \right)  \right)  \right\}  \text{\textrm{d}}u_{a}\text{d}%
u_{b}\text{\textrm{d}}u_{a}^{\prime}\text{d}u_{b}^{\prime}.
\end{multline}
$\mathcal{F}\left[  x,x^{\prime}\right]  $ contains a double path integral as
indicated by the braces, and it can be regarded as an influence
functional~\cite{FeyVer} since it describes the full influence of subsystem
$u$ on subsystem $x$.

The relation~(\ref{eq:Wignerdefinition}) readily allows to write the influence
functional in terms of the initial Wigner distribution function of the $u$
system. Expressed in the center-of-mass and relative coordinate system it
becomes:%
\begin{multline}
\mathcal{F}\left[  x,x^{\prime}\right]  \mathcal{=}\iint\iiiint f\left(
u_{a},p_{a},t_{a}\right)  e^{i\left(  p_{a}\eta_{a}-p_{b}\eta_{b}\right)
/\hbar}\times\label{eq:influencefunc_f}\\
\times\int_{u(t_{a})=u_{a}}^{u(t_{b})=u_{b}}\int_{\eta(t_{a})=\eta_{a}}%
^{\eta(t_{b})=\eta_{b}}\mathcal{D}u\mathcal{D}\eta\exp\left(  \frac{i}{\hbar
}\left(
\begin{array}
[c]{c}%
S_{u}\left[  u+\frac{\eta}{2}\right]  +S_{I}\left[  x,u\mathbf{+}\frac{\eta
}{2}\right] \\
-S_{u}\left[  u\mathbf{-}\frac{\eta}{2}\right]  -S_{I}\left[  x^{\prime
}\mathbf{,}u\mathbf{-}\frac{\eta}{2}\right]
\end{array}
\right)  \right)  \frac{\mathrm{d}p_{a}\mathrm{d}p_{b}}{2\pi\hbar
}\text{\textrm{d}}u_{a}\text{\textrm{d}}\eta_{a}\text{d}u_{b}\text{\textrm{d}%
}\eta_{b},
\end{multline}
where a factor $\delta(\eta_{b})$ was replaced by its plane wave representation.

Proceeding along the lines~(\ref{eq:Probability_nbna}%
--\ref{eq:densitymatrixpropdef}) as in section~\ref{sec:Propagator} one
obtains the time evolution of the \textit{reduced density matrix} of subsystem
$x$:%
\begin{equation}
\rho\left(  x_{b},x_{b}^{\prime},t_{b}\right)  =%
{\displaystyle\iint}
\left\{  \int_{x(t_{a})=x_{a}}^{x(t_{b})=x_{b}}\int_{x^{\prime}(t_{a}%
)=x_{a}^{\prime}}^{x^{\prime}(t_{b})=x_{b}^{\prime}}\mathcal{D}x\mathcal{D}%
x^{\prime}\exp\left(  \frac{i}{\hbar}\left(  S_{x}\left[  x\right]
-S_{x}\left[  x^{\prime}\right]  \right)  \right)  \mathcal{F}\left[
x,x^{\prime}\right]  \right\}  \rho\left(  x_{a},x_{a}^{\prime},t_{a}\right)
\text{\textrm{d}}x_{a}\text{\textrm{d}}x_{a}^{\prime},
\label{eq:Reduceddensitymatrix}%
\end{equation}
which differs from (\ref{eq:densitymatrixpropdef}) merely by the occurrence of
the influence functional in the path integral.

Note that we have assumed the two subsystems to be initially independent so
that the probability $P(j_{a},t_{a})$ of finding $u$ in state $\left\vert
j_{a}\right\rangle $ is independent of the state of $x.$ This means that the
initial total density matrix was supposed to be separable.

Some useful properties of the influence functional are listed below. The
identity%
\begin{equation}
\mathcal{F}^{\ast}\left[  x,x^{\prime}\right]  \mathcal{=F}\left[  x^{\prime
},x\right]  , \label{eq:Fproperty1}%
\end{equation}
follows directly from Eq.~(\ref{eq:influencefunc}) by interchanging $x$ and
$x^{\prime}.$ Note that this property will later ensure that the propagator
for the Wigner function is always a real quantity. Furthermore, if there are a
number of statistically and dynamically independent subsystems $u_{j}$ acting
on $x$ and if $\mathcal{F}^{j}\mathcal{(}x,x^{\prime}\mathcal{)}$ is the
influence functional of the $j$th subsystem on $x\mathbf{,}$ then the total
influence function is the product of all the individual functionals
$\mathcal{F}^{j}:$%
\begin{equation}
\mathcal{F}\left[  x,x^{\prime}\right]  \mathcal{=}%
{\displaystyle\prod\limits_{j=1}^{N}}
\mathcal{F}^{j}\left[  x,x^{\prime}\right]  . \label{eq:Fproperty2}%
\end{equation}
This property is a direct consequence of the total initial density emerging as
a simple product when all $u_{j}$ are statistically independent. If they are
also dynamically independent, then each density matrix can propagate
separately. Finally, it is often convenient to write the influence functional
in the form%
\begin{equation}
\mathcal{F}\left[  x,x^{\prime}\right]  =\exp\left(  \frac{i}{\hbar}%
\Phi\left[  x,x^{\prime}\right]  \right)  \label{eq:Fproperty3}%
\end{equation}
where $\Phi\left[  x,x^{\prime}\right]  $ is called the influence phase. For
independent subsystems, as in~(\ref{eq:Fproperty2})$.$ the corresponding
influence phases add. According to Eq.~(\ref{eq:Fproperty1}), the influence
phase turns out to be antisymmetric under the exchange of $x$ and $x^{\prime}$
if the phase is real, and symmetric if the phase is imaginary. More properties
on influence functionals can be found in \cite{FeyVer}.

It follows from Eq.~(\ref{eq:Reduceddensitymatrix}), and in analogy with the
analysis presented in section \ref{sec:Propagator} until
Eq.~(\ref{eq:PropWignerGeneral}), that the propagator for the reduced Wigner
distribution function becomes%
\begin{equation}
K_{w}\left(  x_{b},p_{b},t_{b}|x_{a},p_{a},t_{a}\right)  =%
{\displaystyle\iint}
\int_{x(t_{a})=x_{a}}^{x(t_{b})=x_{b}}\int_{\xi(t_{a})=\xi_{a}}^{\xi
(t_{b})=\xi_{b}}\mathcal{D}x\mathcal{D}\xi\exp\left(  \frac{i}{\hbar}\left(
\begin{array}
[c]{c}%
-p\xi|_{t_{a}}^{t_{b}}\\
+S_{x}\left[  x\mathbf{+}\frac{\xi}{2}\right] \\
-S_{x}\left[  x-\frac{\xi}{2}\right]
\end{array}
\right)  \right)  \mathcal{F}\left[  x\mathbf{+}\frac{\xi}{2}\mathbf{,}%
x-\frac{\xi}{2}\right]  \frac{\mathrm{d}\xi_{b}\mathrm{d}\xi_{a}}{2\pi\hbar}.
\end{equation}
If the action of the system $x$ is of the form
\begin{equation}
S_{x}\left[  x\right]  =\int_{t_{a}}^{t_{b}}\left(  \frac{m}{2}\dot{x}%
^{2}-V(x)\right)  \mathrm{d}t,
\end{equation}
(i.e., without a magnetic field) the further analysis in
section~\ref{sec:Propagator} until Eq.~(\ref{eq:PropWignerPotential})
simplifies this propagator:
\begin{multline}
K_{w}\left(  x_{b},p_{b},t_{b}|x_{a},p_{a},t_{a}\right)  =\frac{1}{2\pi\hbar
}\int_{\substack{x(t_{a})=x_{a}\\p(t_{a})=m\dot{x}_{a}}}^{\substack{x(t_{b}%
)=x_{b}\\p(t_{b})=m\dot{x}_{b}}}\mathcal{D}x\int\mathcal{D}\xi\mathcal{F}%
\left[  x+\frac{\xi}{2}\mathbf{,}x-\frac{\xi}{2}\right]  \times
\label{eq:PropagatorReducedWigner}\\
\times\exp\left(  -\frac{i}{\hbar}\int_{t_{a}}^{t_{b}}\left(  m\ddot{x}%
\xi\mathbf{+}\left(  V\left(  x\mathbf{+}\frac{\xi}{2}\right)  -V\left(
x\mathbf{-}\frac{\xi}{2}\right)  \right)  \right)  \mathrm{d}t\right)  .
\end{multline}
The crucial ingredient for further development is the influence functional
$\mathcal{F}\left[  x,x^{\prime}\right]  .$ Clearly representing the
propagator of subsystem $u$ under the influence of subsystem $x\mathbf{,}$ the
path integrals in~(\ref{eq:influencefunc}) can be calculated analytically if
$S_{u}\left[  u\right]  +S_{I}\left[  x\mathbf{,}u\right]  $ is quadratic in
$u\mathbf{.}$ Below the results for a bare harmonic action $S_{u}$ with a
linear coupling $S_{I}$ will be discussed in some detail.

\section{Harmonic subsystems with linear coupling\label{sec:Harmonicinfluence}%
}

\subsection{A single oscillator}

In this section we have reduced the subsystem $u$ to a single harmonic
oscillator with mass $M$ and with potential energy $V\left(  u\right)
=M\omega^{2}u^{2}/2$, interacting with the subsystem $x$ of interest. Taking
the interaction energy to be $u\cdot\gamma(x)$, with an arbitrary function
$\gamma(x)$ and a linear dependence on $u$, we are left with $S_{I}\left[
x,u\right]  =-\int_{t_{a}}^{t_{b}}u\gamma_{t}(x)\mathrm{d}t$, where we added a
subscript $t$ to $\gamma$ to remember the time at which its path variable $x$
should be evaluated. The total action of subsystem $u\mathbf{,}$ to be used in
the influence functional~(\ref{eq:influencefunc_f}), is thus%
\begin{equation}
S_{u}\left[  u\right]  +S_{I}\left[  x\mathbf{,}u\right]  =\int_{t_{a}}%
^{t_{b}}\left(  \frac{M}{2}\dot{u}^{2}-\frac{M}{2}\omega^{2}u^{2}-u\gamma
_{t}(x)\right)  \mathrm{d}t.
\end{equation}
The argument in the exponent of Eq.~(\ref{eq:influencefunc_f}) thus becomes
linear in $\eta$ and in $u\mathbf{:}$%
\begin{multline*}
\mathcal{F}\left[  x,x^{\prime}\right]  \mathcal{=}\iint\iiiint f\left(
u_{a},p_{a},t_{a}\right)  e^{i\left(  p_{a}\eta_{a}-p_{b}\eta_{b}\right)
/\hbar}\int_{u(t_{a})=u_{a}}^{u(t_{b})=u_{b}}\mathcal{D}u\exp\left(  -\frac
{i}{\hbar}\int_{t_{a}}^{t_{b}}u\left(  \gamma_{t}(x)-\gamma_{t}(x^{\prime
})\right)  \mathrm{d}t\right)  \times\\
\times\int_{\eta(t_{a})=\eta_{a}}^{\eta(t_{b})=\eta_{b}}\mathcal{D}\eta
\exp\left(  -\frac{i}{\hbar}\int_{t_{a}}^{t_{b}}\left(  M\dot{u}\dot{\eta
}-M\omega^{2}u\eta\mathbf{-}\eta\frac{\gamma_{t}(x)+\gamma_{t}(x^{\prime})}%
{2}\right)  \mathrm{d}t\right)  \frac{\mathrm{d}p_{a}\mathrm{d}p_{b}}%
{2\pi\hbar}\text{\textrm{d}}u_{a}\text{\textrm{d}}\eta_{a}\text{d}%
u_{b}\text{\textrm{d}}\eta_{b}.
\end{multline*}
After an integration by parts of the kinetic term $\int_{t_{a}}^{t_{b}}\dot
{u}\dot{\eta}\mathrm{d}t=\dot{u}_{b}\dot{\eta}_{b}-\dot{u}_{a}\dot{\eta}%
_{a}-\mathbf{\int_{t_{a}}^{t_{b}}}\ddot{u}\eta\mathrm{d}t,$ and imposing
$M\dot{u}_{a,b}=p_{a,b},$ the path integral over $\eta$ becomes unconstrained:%
\begin{multline}
\mathcal{F}\left[  x,x^{\prime}\right]  =\iiiint f\left(  u_{a},p_{a}%
,t_{a}\right)  \int_{\substack{u(t_{a})=u_{a}\\p(t_{a})=Mu_{a}}%
}^{\substack{u(t_{b})=u_{b}\\p(t_{b})=Mu_{b}}}\mathcal{D}u\exp\left(
-\frac{i}{\hbar}\int_{t_{a}}^{t_{b}}\left(  \gamma_{t}(x)-\gamma_{t}%
(x^{\prime})\right)  u\mathrm{d}t\right)  \times\label{eq:phasespaceF1osc}\\
\times\int\mathcal{D}\eta\exp\left(  \frac{i}{\hbar}\int_{t_{a}}^{t_{b}%
}\left(  -M\ddot{u}-M\omega^{2}u\mathbf{-}\frac{\gamma_{t}(x^{\prime}%
)+\gamma_{t}(x)}{2}\right)  \eta\mathrm{d}t\right)  \frac{\mathrm{d}%
p_{a}\mathrm{d}p_{b}\text{\textrm{d}}u_{a}\text{d}u_{b}}{2\pi\hbar}.
\end{multline}
The path integral over all $\eta$ in the last line\textbf{ }is of the form of
the quadratic path integral~(\ref{eq:Harmonicexample}), and restricts the
phase space trajectories of $u$%
\begin{equation}
M\left(  \ddot{u}+\omega^{2}u\right)  \mathbf{+}\frac{\gamma_{t}%
(x)\mathbf{+}\gamma_{t}(x^{\prime})}{2}=0,
\end{equation}
with the formal solution%
\[
u\left(  t\right)  =u_{a}\cos\omega\left(  t-t_{a}\right)  +\frac{\dot{u}_{a}%
}{\omega}\sin\omega\left(  t-t_{a}\right)  -\int_{t_{a}}^{t}\frac{\gamma
_{s}(x)+\gamma_{s}(x^{\prime})}{2M}\frac{\sin\omega\left(  t-s\right)
}{\omega}\,\text{\textrm{d}}s.
\]
The remaining path integral on the first line of~(\ref{eq:phasespaceF1osc})
imposes that the initial velocity is $\dot{u}_{a}=p_{a}/M.$ The trajectories
in $u$ are thus reduced to a single path, with the conditions $p_{a,b}%
=M\dot{u}_{a,b}$ at the end points, which also eliminate the integrations over
$u_{b}$ and $p_{b}.$ One thus readily arrives at
\begin{multline}
\mathcal{F}\left[  x,x^{\prime}\right]  =\exp\left(  \frac{i}{2\hbar M\omega
}\int_{t_{a}}^{t_{b}}\int_{t_{a}}^{t}\left(  \gamma_{s}(x)+\gamma
_{s}(x^{\prime})\right)  \left(  \gamma_{t}(x)-\gamma_{t}(x^{\prime})\right)
\sin\omega\left(  t-s\right)  \,\text{\textrm{d}}s\mathrm{d}t\right)
\times\label{eq:Influencegeneralgamma}\\
\times\iint f\left(  u_{a},p_{a},t_{a}\right)  \exp\left(  -\frac{i}{\hbar
}\int_{t_{a}}^{t_{b}}\left(  u_{a}\cos\omega\left(  t-t_{a}\right)
+\frac{p_{a}}{M\omega}\sin\omega\left(  t-t_{a}\right)  \right)  \left(
\gamma_{t}(x)-\gamma_{t}(x^{\prime})\right)  \mathrm{d}t\right)
\text{\textrm{d}}r_{a}\text{\textrm{d}}p_{a},
\end{multline}
One might be concerned about the normalizing factors accompanying the several
delta functions in the derivation, but this possible problem is resolved by
considering the uncoupled limit $\gamma=0.$ An alternative derivation, using
the explicit solution of the Feynman path integral~\cite{cit:FeynmanHibbs1965}
for the action $\left(  S_{u}\left[  u\right]  +S_{I}\left[  x,u\right]
\right)  ,$ confirms this result.

Clearly the first line in expression (\ref{eq:Influencegeneralgamma}) is
independent of the initial state of the harmonic oscillator. This term
represents an effective interaction of system $x$ with itself. The second line
is an expectation value which transfers all necessary information about the
initial state of the harmonic $u$ system into $x\mathbf{.}$

\subsubsection{Example: initial wave packet}

Despite the classical trajectories which govern its dynamics, the influence
functional~(\ref{eq:Influencegeneralgamma}) is intrinsically of quantum
mechanical nature, because the initial Wigner distribution function $f\left(
u_{a},p_{a},t_{a}\right)  $ of the oscillator is bound to satisfy the
uncertainty principle. A sharply defined initial distribution like
$\delta(u_{a}-u_{0})\delta(p_{a}-p_{0})$ can not be of the
form~(\ref{eq:Wignerdefinition}). However, a valid initial wave function could
be a Gaussian wave packet%
\[
\Psi_{G}\left(  u,t_{a}\right)  =\frac{1}{\sqrt{\Delta\sqrt{2\pi}}}\exp\left(
-\frac{\left(  u-u_{0}\right)  ^{2}}{4\Delta^{2}}\right)  e^{ip_{0}%
u\mathbf{/\hbar}}.
\]
From the corresponding density matrix $\rho_{G}\left(  u\mathbf{,}u^{\prime
},t\right)  =\Psi_{G}^{\ast}\left(  u^{\prime},t\right)  \Psi_{G}\left(
u,t\right)  $ one easily finds the Wigner distribution
function~(\ref{eq:Wignerdefinition}) of this wave packet%
\begin{equation}
f_{G}\left(  u,p,t_{a}\right)  =\frac{1}{\pi\hbar}\exp\left(  -\frac{\left(
u-u_{0}\right)  ^{2}}{2\Delta^{2}}-2\Delta^{2}\left(  \frac{p-p_{0}}{\hbar
}\right)  ^{2}\right)  .
\end{equation}
Then, by performing the integrations in Eq.(\ref{eq:Influencegeneralgamma})
and rearranging some terms one finds the following influence
phase~(\ref{eq:Fproperty3}):%
\begin{multline}
\Phi_{G}\left[  x,x^{\prime}\right]  =\frac{1}{2M\omega}\int_{t_{a}}^{t_{b}%
}\int_{t_{a}}^{t}\left(  \gamma_{s}(x)+\gamma_{s}(x^{\prime})\right)  \left(
\gamma_{t}(x)-\gamma_{t}(x^{\prime})\right)  \sin\omega\left(  t-s\right)
\,\text{\textrm{d}}s\mathrm{d}t\\
-u_{0}\int_{t_{a}}^{t_{b}}\left(  \gamma_{t}(x)-\gamma_{t}(x^{\prime})\right)
\cos\omega\left(  t-t_{a}\right)  \mathrm{d}t-\frac{p_{0}}{M\omega}\int
_{t_{a}}^{t_{b}}\left(  \gamma_{t}(x)-\gamma_{t}(x^{\prime})\right)
\sin\omega\left(  t-t_{a}\right)  \mathrm{d}t\\
+\frac{i}{4\hbar}\int_{t_{a}}^{t_{b}}\int_{t_{a}}^{t_{b}}\left(  \gamma
_{s}(x)-\gamma_{s}(x^{\prime})\right)  \left(  \gamma_{t}(x)-\gamma
_{t}(x^{\prime})\right)  \left(
\begin{array}
[c]{c}%
\left(  \Delta^{2}+\frac{\hbar^{2}}{4\Delta^{2}M^{2}\omega^{2}}\right)
\cos\omega\left(  s-t\right) \\
+\left(  \Delta^{2}-\frac{\hbar^{2}}{4\Delta^{2}M^{2}\omega^{2}}\right)
\cos\omega\left(  s+t-2t_{a}\right)
\end{array}
\right)  \mathrm{d}s\mathrm{d}t
\end{multline}
The real part of the influence phase is the same as one would obtain from a
(forbidden) initial Wigner distribution function $\delta(u_{a}-u_{0}%
)\delta(p_{a}-p_{0})$. Besides the effective interacting of system $x$ with
itself, the real part now contains an external driving potential which
oscillates in time with a frequency $\omega$ and its spatial dependence is
given by $\gamma(x)$. The magnitude of this driving potential depends on the
initial average (vacuum) displacement of the oscillator $\left(  u_{0}%
,p_{0}\right)  .$ The imaginary part of the influence phase results from the
uncertainty on the initial position and momentum of $u$; its physical
significance becomes more apparent in the next example. Note that some terms
vanish under specific conditions, e.g., if the average initial position or
momentum are zero. The last term becomes zero whenever $\Delta^{2}=\frac
{\hbar}{2M\omega}.$ This condition is satisfied if the oscillator was
initially in an unsqueezed coherent state. In that case one finds the ground
state or vacuum influence phase:%
\begin{multline}
\Phi_{vac}\left[  x,x^{\prime}\right]  =\frac{1}{2M\omega}\int_{t_{a}}^{t_{b}%
}\int_{t_{a}}^{t}\left(  \gamma_{s}(x)+\gamma_{s}(x^{\prime})\right)  \left(
\gamma_{t}(x)-\gamma_{t}(x^{\prime})\right)  \sin\omega\left(  t-s\right)
\,\text{\textrm{d}}s\mathrm{d}t\\
+\frac{i}{4M\omega}\int_{t_{a}}^{t_{b}}\int_{t_{a}}^{t_{b}}\left(  \gamma
_{s}(x)-\gamma_{s}(x^{\prime})\right)  \left(  \gamma_{t}(x)-\gamma
_{t}(x^{\prime})\right)  \cos\omega\left(  s-t\right)  \mathrm{d}s\mathrm{d}t.
\end{multline}

\subsubsection{Example: thermal equilibrium}

The case of thermal equilibrium at the start of course deserves some
additional attention. The initial equilibrium Wigner function~\cite{BMNewton}
of the harmonic oscillator $u$ is then given by%
\begin{equation}
f_{eq}\left(  u_{a},p_{a},t_{a}\right)  =\frac{\tanh\frac{1}{2}\beta
\hbar\omega}{\pi\hbar}\exp\left(  -\frac{\tanh\frac{1}{2}\beta\hbar\omega
}{\hbar\omega}\left(  M\omega^{2}u_{a}^{2}+\frac{p_{a}^{2}}{M}\right)
\right)  . \label{eq:Wignerequilibrium}%
\end{equation}
It follows from Eq.~(\ref{eq:Influencegeneralgamma}) that the equilibrium
influence phase $\Phi_{eq}\left[  x,x^{\prime}\right]  $ is%
\begin{multline}
\Phi_{eq}\left[  x,x^{\prime}\right]  =\frac{1}{2M\omega}\int_{t_{a}}^{t_{b}%
}\int_{t_{a}}^{t}\left(  \gamma_{s}(x)+\gamma_{s}(x^{\prime})\right)  \left(
\gamma_{t}(x)-\gamma_{t}(x^{\prime})\right)  \sin\omega\left(  t-s\right)
\,\text{\textrm{d}}s\mathrm{d}t+\\
+\frac{i\coth\frac{1}{2}\beta\hbar\omega}{4M\omega}\int_{t_{a}}^{t_{b}}%
\int_{t_{a}}^{t_{b}}\left(  \gamma_{t}(x)-\gamma_{t}(x^{\prime})\right)
\left(  \gamma_{s}(x)-\gamma_{s}(x^{\prime})\right)  \cos\omega\left(
t-s\right)  \mathrm{d}s\mathrm{d}t
\end{multline}
While the real parts of $\Phi_{eq}$ and $\Phi_{vac}$ are the same, the
imaginary part of $\Phi_{eq}$ has increased by a factor $\coth\left(
\frac{\beta\hbar\omega}{2}\right)  $ as compared to $\Phi_{vac},$ as a
consequence of thermal broadening of the distribution. Note now that
$\coth\left(  \frac{\beta\hbar\omega}{2}\right)  =\left[  1+n_{B}%
(\omega)\right]  +n_{B}(\omega),$ with $n_{B}(\omega)$ the Bose-Einstein
distribution with zero chemical potential. From a physics point of view one
would therefore associate the imaginary part of $\Phi_{eq}$ with the emission
and absorption of the quanta represented by the oscillator. When the
temperature is zero system $x$ can only interact with the zero point
fluctuations of $u$, allowing only losses trough spontaneous emission. A more
detailed discussion on this can be found in \cite{FeyVer}.

\subsection{Many independent oscillators}

The generalization of the result from one to many oscillators is trivial if
the oscillators are independent. According to property~(\ref{eq:Fproperty2})
the total influence functional then becomes the product of all the individual
influence functionals. Consider every oscillator $u_{j}$ to have a mass
$M_{j}$, a frequency $\omega_{j}$ and consider the interaction energy with $x$
to be $u_{j}\cdot\gamma_{j}(x).$ Then we immediately arrive at the following
expression for the total influence phase of a collection of $N$ oscillators%
\begin{multline}
\mathcal{F}_{N}\left[  x,x^{\prime}\right]  =\exp\left(  \frac{i}{\hbar}%
\sum_{j=0}^{N}\frac{1}{2M_{j}\omega_{j}}\int_{t_{a}}^{t_{b}}\int_{t_{a}}%
^{t}\left(  \gamma_{j}(x_{s})+\gamma_{j}(x_{s}^{\prime})\right)  \left(
\gamma_{j}(x_{t})-\gamma_{j}(x_{t}^{\prime})\right)  \sin\omega_{j}\left(
t-s\right)  \,\text{\textrm{d}}s\mathrm{d}t\right)  \times
\label{eq:Influencefunctmany}\\
\times\prod_{j=0}^{N}\iint f_{j}\left(  u_{a},p_{a},t_{a}\right)  \exp\left(
-\frac{i}{\hbar}\int_{t_{a}}^{t_{b}}\left(  u_{a}\cos\omega_{j}\left(
t-t_{a}\right)  +\frac{p_{a}}{M_{j}\omega_{j}}\sin\omega_{j}\left(
t-t_{a}\right)  \right)  \left(  \gamma_{j}(x_{t})-\gamma_{j}(x_{t}^{\prime
})\right)  \mathrm{d}t\right)  \text{\textrm{d}}r_{a}\text{\textrm{d}}p_{a},
\end{multline}
where $f_{j}\left(  u_{a},p_{a},t_{a}\right)  $ represents the initial Wigner
function of the $j$th oscillator$.$ Let us consider a simple example now that
can easily be generalized to more complicated situations.

\subsubsection{Example: Equilibrium oscillators with bilinear coupling}

If we assume the interaction energy to be bilinear in $\left\{  u_{j}%
,x\right\}  ,$ such that $u\cdot\gamma_{j}(x)=\gamma_{j}u\cdot x,$ and if
additionally all oscillators are initially in thermal equilibrium
(\ref{eq:Wignerequilibrium}), then the influence phase associated with
influence functional (\ref{eq:Influencefunctmany}) is given by%
\begin{multline*}
\Phi_{eq,N}\left[  x,x^{\prime}\right]  =\sum_{j=0}^{N}\frac{\gamma_{j}^{2}%
}{2M_{j}\omega_{j}}\int_{t_{a}}^{t_{b}}\int_{t_{a}}^{t}\left(  x_{s}%
+x_{s}^{\prime}\right)  \left(  x_{t}-x_{t}^{\prime}\right)  \sin\omega
_{j}\left(  t-s\right)  \,\text{\textrm{d}}s\mathrm{d}t\\
+\sum_{j=0}^{N}\frac{i\gamma_{j}^{2}}{4M_{j}\omega_{j}}\coth\frac{1}{2}%
\beta\hbar\omega_{j}\int_{t_{a}}^{t_{b}}\int_{t_{a}}^{t_{b}}\left(
x_{t}-x_{t}^{\prime})\right)  \left(  x_{s}-x_{s}^{\prime})\right)  \cos
\omega_{j}\left(  t-s\right)  \mathrm{d}s\mathrm{d}t.
\end{multline*}
In the continuum limit, when $N\rightarrow\infty$ while $\omega_{j+1}%
-\omega_{j}\rightarrow0,$ we can assume there is a distribution of
oscillators, such that the relevant weight $\Gamma(\omega)$\textrm{d}$\omega$
of the oscillators between $\omega$ and $\omega+$\textrm{d}$\omega$ is%
\[
\Gamma(\omega)=\sum_{j=0}^{\infty}\frac{\gamma_{j}^{2}}{M_{j}\omega_{j}}%
\delta(\omega-\omega_{j}).
\]
In this case the influence phase becomes%
\begin{multline*}
\Phi_{eq,many}\left[  x,x^{\prime}\right]  =\frac{1}{2}\int_{t_{a}}^{t_{b}%
}\int_{t_{a}}^{t}\left(  \int_{0}^{\infty}\Gamma(\omega)\sin\omega\left(
t-s\right)  \,\mathrm{d}\omega\right)  \left(  x_{s}+x_{s}^{\prime}\right)
\left(  x_{t}-x_{t}^{\prime}\right)  \text{\textrm{d}}s\mathrm{d}t\\
+\frac{i}{\hbar}\int_{t_{a}}^{t_{b}}\int_{t_{a}}^{t_{b}}\left[  \int
_{0}^{\infty}\left(  \frac{\hbar\Gamma(\omega)}{4}\coth\frac{\beta\hbar\omega
}{2}\right)  \cos\omega\left(  t-s\right)  \mathrm{d}\omega\right]  \left(
x_{t}-x_{t}^{\prime})\right)  \left(  x_{s}-x_{s}^{\prime})\right)
\mathrm{d}s\mathrm{d}t.
\end{multline*}
If $x$ is the coordinate of an otherwise free particle with mass $m$ we find
the following propagator for its reduced Wigner function%
\begin{multline*}
K_{w}\left(  x_{b},p_{b},t_{b}|x_{a},p_{a},t_{a}\right)  =\frac{1}{2\pi\hbar
}\int_{\substack{x(t_{a})=x_{a}\\p(t_{a})=m\dot{x}_{a}}}^{\substack{x(t_{b}%
)=x_{b}\\p(t_{b})=m\dot{x}_{b}}}\mathcal{D}x\int\mathcal{D}\xi\exp\left(
-\frac{1}{\hbar^{2}}\int_{t_{a}}^{t_{b}}\int_{t_{a}}^{t_{b}}R(t-s)\xi_{t}%
\cdot\xi_{s}\mathrm{d}s\mathrm{d}t\right) \\
\times\exp\left(  -\frac{i}{\hbar}\int_{t_{a}}^{t_{b}}\left(  m\ddot{x}%
-\int_{t_{a}}^{t}A(t-s)x_{s}\mathrm{d}s\right)  \cdot\xi_{t}\mathrm{d}%
t\right)  ,
\end{multline*}
with $A(t)$ and $R(t)$ defined as:%
\begin{align*}
A(t)  &  =\int_{0}^{\infty}\Gamma(\omega)\sin(\omega t)\text{\textrm{d}}%
\omega,\\
R(t)  &  =\int_{0}^{\infty}\text{ }\frac{\hbar\Gamma(\omega)}{4}\coth\left(
\frac{\beta\hbar\omega}{2}\right)  \cos(\omega t)\text{\textrm{d}}\omega.
\end{align*}

\subsubsection{Example: Caldeira-Legett model and thermalization}

In the specific case that $\Gamma(\omega)=2\eta\omega$, as considered in
detail by Caldeira and Leggett in \cite{CaldeiraLeggett}, we obtain
$A(t)=-2\eta\delta^{\prime}(t).$ If additionally the temperature $T_{b}$ is
high enough such that we can approximate $\left[  \frac{\hbar\eta\omega}%
{2}\coth\left(  \frac{\beta\hbar\omega}{2}\right)  \right]  \approx\frac{\eta
}{\beta}+\mathcal{O}(\beta),$ we find $R(t)=\eta kT_{b}\delta(t-s).$ Therefore
we arrive\footnote{under some debatebly
assumptions\cite{CaldeiraLeggett,Rousenau} about the limits in the integral.}
at the following expression for the propagator in the high temperature limit%
\[
K_{w}\left(  x_{b},p_{b},t_{b}|x_{a},p_{a},t_{a}\right)  =\frac{1}{2\pi\hbar
}\int_{\substack{x(t_{a})=x_{a}\\p(t_{a})=m\dot{x}_{a}}}^{\substack{x(t_{b}%
)=x_{b}\\p(t_{b})=m\dot{x}_{b}}}\mathcal{D}x\int\mathcal{D}\xi\exp\left(
-\frac{\eta kT_{b}}{\hbar^{2}}\int_{t_{a}}^{t_{b}}\xi_{t}^{2}\mathrm{d}%
t\right)  \exp\left(  -\frac{i}{\hbar}\int_{t_{a}}^{t_{b}}\left(  m\ddot
{x}+\eta\dot{x}\right)  \cdot\xi_{t}\mathrm{d}t\right)  .
\]
It can easily be shown, for example by timeslicing the $\xi$ path integral,
that this propagator can be rewritten as a single path integral:%
\[
K_{w}\left(  x_{b},p_{b},t_{b}|x_{a},p_{a},t_{a}\right)  =\int
_{\substack{x(t_{a})=x_{a}\\p(t_{a})=m\dot{x}_{a}}}^{\substack{x(t_{b}%
)=x_{b}\\p(t_{b})=m\dot{x}_{b}}}\mathcal{D}x\exp\left(  -\frac{1}{4\eta
kT_{b}}\int_{t_{a}}^{t_{b}}\left(  m\ddot{x}+\eta\dot{x}\right)
^{2}\mathrm{d}t\right)  .
\]
If we are not interested in the real space motion of $x$, for example because
the initial distribution of the particle is homogeneous in space, but only in
the marginal propagator $K_{w}\left(  p_{b},t_{b}|p_{a},t_{a}\right)  $ to go
from $p_{a}$ to $p_{b}$, then we get%
\[
K_{w}\left(  p_{b},t_{b}|p_{a},t_{a}\right)  =\int_{p(t_{a})=p_{a}}%
^{p(t_{b})=p_{b}}\mathcal{D}p\exp\left(  -\frac{1}{2\eta kT_{b}}\int_{t_{a}%
}^{t_{b}}\left(  \frac{\left(  \dot{p}\right)  ^{2}}{2}+\left(  \frac{\eta}%
{m}\right)  ^{2}\frac{p^{2}}{2}+\frac{\eta}{m}\dot{p}p\right)  \text{d}%
t\right)  .
\]
An integration by parts shows that the term in $\dot{p}p$ only contributes at
the boundaries:%
\[
K_{w}\left(  p_{b},t_{b}|p_{a},t_{a}\right)  =\exp\left(  -\frac{p_{b}%
^{2}-p_{a}^{2}}{4mkT_{b}}\right)  \int_{p(t_{a})=p_{a}}^{p(t_{b})=p_{b}%
}\mathcal{D}p\exp\left(  -\frac{1}{2\eta kT_{b}}\int_{t_{a}}^{t_{b}}\left(
\frac{\left(  \dot{p}\right)  ^{2}}{2}+\left(  \frac{\eta}{m}\right)
^{2}\frac{p^{2}}{2}\right)  \text{d}t\right)  ,
\]
The remaining path integral is a simple Gaussian Feynman path integral. This
path integral can be solved with standard techniques which yield the following
expression for the reduced momentum space Wigner propagator:%
\[
K_{w}\left(  p_{b},t_{b}|p_{a},t_{a}\right)  =\frac{1}{\sqrt{2\pi
mkT_{b}\left(  1-\exp\left(  -\frac{2\eta}{m}\left(  t_{b}-t_{a}\right)
\right)  \right)  }}\exp\left(  -\frac{1}{2mkT_{b}}\frac{\left(  p_{b}%
-p_{a}\exp\left(  -\frac{\eta}{m}\left(  t_{b}-t_{a}\right)  \right)  \right)
^{2}}{1-\exp\left(  -\frac{2\eta}{m}\left(  t_{b}-t_{a}\right)  \right)
}\right)  .
\]
The maximal transition probability is, as expected, attained along the
solution of the classical equation of motion $\dot{p}+\frac{\eta}{m}p=0,$ i.e.
when \ $p_{b}=p_{a}\exp\left(  -\frac{\eta}{m}\left(  t_{b}-t_{a}\right)
\right)  $. The variance of the propagator is given by%
\[
\sigma^{2}=mkT_{b}\left(  1-\exp\left(  -\frac{2\eta}{m}\left(  t_{b}%
-t_{a}\right)  \right)  \right)  .
\]
The time evolution of the effective temperature $T_{e}$ of the particle $x$ is
therefore given by
\[
T_{e}(t_{b}-t_{a})=T_{b}\left(  1-\exp\left(  -\frac{2\eta}{m}\left(
t_{b}-t_{a}\right)  \right)  \right)  ,
\]
such that the system thermalizes at a characteristic time $\tau=\frac{m}%
{2\eta}.$ Finally consider the time $t_{b}-t_{a}\gg\tau,$ such that the system
is thermalized and $T_{e}\approx T_{b}$, then the propagator becomes%
\[
\lim_{\left(  t_{b}-t_{a}\right)  \rightarrow\infty}K_{w}\left(  p_{b}%
,t_{b}|p_{a},t_{a}\right)  =\frac{1}{\sqrt{2\pi mkT_{b}}}\exp\left(  -\frac
{1}{kT_{b}}\frac{p_{b}^{2}}{2m}\right)  .
\]
This means that the reduced Wigner function of the particle thermalizes into a
Maxwell-Boltzmann distribution, regardless its initial Wigner function.

\section{Conclusion}

In conclusion we have used Feynman's Lagrangian description of quantum
mechanics to express the propagator of the Wigner function as a path integral.
Propagating Wigner functions rather than wave functions has a double
advantage. First of all one can directly propagate uncertain initial
configurations in time instead of pure states only. Secondly the attained
Wigner function propagator becomes a delta function in the classical
trajectory for all harmonic problems and subproblems.In contrast to our
previous derivation in \cite{SBMpathintegral} based on the Wigner-Liouville
equation, the present treatment does not rely on a canonical quantization
procedure of the Hamiltonian. This Lagrangian formulation, with the help of
influence functionals, allows for a transparent description of two interacting
subsystems, and to find the reduced Wigner function propagator of one of the
two subsystems. In the last section we have generalized this result to a
system interacting with many other, mutually independent, subsystems. As an
example we considered in more detail the model by Caldeira and Legett, and
showed how to calculate the reduced Wigner function propagator for it, using
the techniques explained in this manuscript.%

\newpage
\appendix{}

\section{Wigner propagator for quadratic potentials}

In this appendix the propagator~(\ref{eq:Harmonicexample}) of the Wigner
function for quadratic potentials of the form%
\begin{equation}
V_{\text{quad}}(x,t)=a(t)+b(t)x+c(t)x^{2} \label{eq:Vquad}%
\end{equation}
is derived from the well known~\cite{cit:Schulman1981} Feynman propagator
$K_{\text{quad}}\left(  x_{b},t_{b}|x_{a},t_{a}\right)  $ for this type of
potentials:%
\begin{align}
K_{\text{quad}}\left(  x_{b},t_{b}|x_{a},t_{a}\right)   &  =\sqrt{\frac
{m}{2\pi i\hbar f\left(  t_{b},t_{a}\right)  }}\exp\left(  \frac{i}{\hbar
}S_{\text{quad}}\left(  x_{b},t_{b}|x_{a},t_{a}\right)  \right)
,\label{eq:K_quadr}\\
S_{\text{quad}}\left(  x_{b},t_{b}|x_{a},t_{a}\right)   &  =\int_{t_{a}%
}^{t_{b}}\left(  \frac{m}{2}\dot{x}^{2}-a\left(  t\right)  -b\left(  t\right)
x-c\left(  t\right)  x^{2}\right)  \mathrm{d}t, \label{eq:S_quadr}%
\end{align}
where $S_{\text{quad}}\left(  x_{b},t_{b}|x_{a},t_{a}\right)  $ is the action
of the system along a classical trajectory from $\left(  x_{a},t_{a}\right)  $
to $\left(  x_{b},t_{b}\right)  ,$ to be determined from the classical
equation of motion%
\begin{equation}
m\ddot{x}+2c\left(  t\right)  x=-b\left(  t\right)  \text{ with }%
\begin{array}
[c]{l}%
x\left(  t_{a}\right)  =x_{a},\\
x\left(  t_{b}\right)  =x_{b}.
\end{array}
\label{eq:eqmotionx_quad}%
\end{equation}
The function $f\left(  t_{b},t_{a}\right)  $ only depends on the initial and
final time, and is independent of the positions and momenta. It is the
solution of the differential equation%
\begin{equation}
\left(  m\frac{d^{2}}{dt^{2}}+2c\left(  t\right)  \right)  f\left(
t,t_{a}\right)  =0\text{ with }%
\begin{array}
[c]{l}%
f\left(  t_{a},t_{a}\right)  =0,\\
\left.  \frac{d}{dt}f\left(  t,t_{a}\right)  \right\vert _{t=t_{a}}=1.
\end{array}
\label{eq:eqmotionf_quad}%
\end{equation}
For general time dependence of $c\left(  t\right)  ,$ these differential
equations rarely have a solution in closed form, but it is sure that two
linearly independent solutions, say $x_{1}\left(  t\right)  $ and
$x_{2}\left(  t\right)  ,$ of the homogeneous equations exist:%
\begin{equation}
m\ddot{x}_{1,2}+2c\left(  t\right)  x_{1,2}=0. \label{eq:x12diffvgl}%
\end{equation}
Their Wronskiaan $\frac{dx_{1}\left(  t\right)  }{dt}x_{2}\left(  t\right)
-\frac{dx_{2}\left(  t\right)  }{dt}x_{1}\left(  t\right)  $ is independent of
$t,$ because~(\ref{eq:x12diffvgl}) reveals that its time derivative is zero%
\begin{equation}
\frac{dx_{1}\left(  t\right)  }{dt}x_{2}\left(  t\right)  -\frac{dx_{2}\left(
t\right)  }{dt}x_{1}\left(  t\right)  =W\text{ independent of }t. \label{eq:W}%
\end{equation}
Since $f\left(  t,t_{a}\right)  $ is also a solution of the \emph{homogeneous}
differential equation~(\ref{eq:x12diffvgl}), it is a linear combination of
$x_{1}\left(  t\right)  $ and $x_{2}\left(  t\right)  .$ Taking the boundary
conditions into account it becomes%
\begin{equation}
f\left(  t,t_{a}\right)  =\frac{h\left(  t,t_{a}\right)  }{W}\text{ with
}h\left(  s,t\right)  =x_{1}\left(  s\right)  x_{2}\left(  t\right)
-x_{2}\left(  s\right)  x_{1}\left(  t\right)  .
\end{equation}
$\allowbreak$

If one imposes that the solution of the homogeneous differential
equation~(\ref{eq:x12diffvgl}) exhausts the boundary conditions at $t_{a}$ and
$t_{b},$ the trajectory $x\left(  t\right)  $ is of the form%
\begin{equation}
x\left(  t\right)  =\frac{h\left(  t,t_{b}\right)  }{h\left(  t_{a}%
,t_{b}\right)  }x_{a}+\frac{h\left(  t_{a},t\right)  }{h\left(  t_{a}%
,t_{b}\right)  }x_{b}+x_{p}\left(  t\right)  , \label{eq:x(t)quadr}%
\end{equation}
where the particular solution $x_{p}\left(  t\right)  $ has to satisfy the
boundary conditions $x_{p}\left(  t_{a}\right)  =0=x_{p}\left(  t_{b}\right)
.$ It is easily found by the variation of parameters method, with the result:%
\begin{equation}
x_{p}\left(  t\right)  =-\frac{h\left(  t,t_{b}\right)  }{h\left(  t_{a}%
,t_{b}\right)  }\int_{t_{a}}^{t}\frac{b\left(  s\right)  }{m}\frac{h\left(
t_{a},s\right)  }{W}\,\mathrm{d}s-\frac{h\left(  t_{a},t\right)  }{h\left(
t_{a},t_{b}\right)  }\int_{t}^{t_{b}}\frac{b\left(  s\right)  }{m}%
\frac{h\left(  s,t_{b}\right)  }{W}\,\mathrm{d}s. \label{eq:xp(t)quadr}%
\end{equation}
It is fairly easy to calculate the initial and final velocities $\dot{x}%
_{a,b}$, which are of particular relevance below:%
\begin{align}
\dot{x}_{a}  &  =\frac{1}{h\left(  t_{a},t_{b}\right)  }\left(  x_{a}%
\frac{\partial h\left(  t_{a},t_{b}\right)  }{\partial t_{a}}-x_{b}%
W+\int_{t_{a}}^{t_{b}}\frac{b\left(  s\right)  }{m}\,h\left(  s,t_{b}\right)
\mathrm{d}s\right)  ,\label{eq:xdot(t)quadr_ta}\\
\dot{x}_{b}  &  =\frac{1}{h\left(  t_{a},t_{b}\right)  }\left(  x_{a}%
W+x_{b}\frac{\partial h\left(  t_{a},t_{b}\right)  }{\partial t_{b}}%
-\int_{t_{a}}^{t_{b}}\frac{b\left(  s\right)  }{m}h\left(  t_{a},s\right)
\,\mathrm{d}s\right)  . \label{eq:xdot(t)quadr_tb}%
\end{align}
Applying an integration by parts $\int\dot{x}^{2}\mathrm{d}t=x\dot{x}-\int
x\ddot{x}\mathrm{d}t$ in the kinetic contribution to the classical action, it
can be rewritten as
\begin{multline}
S_{\text{quad}}\left(  x_{b},t_{b}|x_{a},t_{a}\right)  =\frac{m}{2}\left(
\left[  x\dot{x}\right]  _{t=t_{a}}^{t=t_{b}}-\int_{t_{a}}^{t_{b}}%
\frac{b\left(  t\right)  }{m}x\left(  t\right)  \mathrm{d}t\right) \\
=\frac{m}{h\left(  t_{a},t_{b}\right)  }\left(  \frac{x_{b}^{2}\frac{dh\left(
t_{a},t_{b}\right)  }{dt_{b}}-x_{a}^{2}\frac{dh\left(  t_{a},t_{b}\right)
}{dt_{a}}}{2}+Wx_{a}x_{b}-x_{b}\int_{t_{a}}^{t_{b}}\frac{b\left(  s\right)
}{m}h\left(  t_{a},s\right)  \,\mathrm{d}s-x_{a}\int_{t_{a}}^{t_{b}}%
\frac{b\left(  s\right)  }{m}h\left(  s,t_{b}\right)  \,\mathrm{d}s\right)
-\\
-\frac{1}{2}\int_{t_{a}}^{t_{b}}b\left(  s\right)  x_{p}\left(  s\right)
\,\mathrm{d}s,
\end{multline}
where the boundary velocities and the homogenous contribution $x\left(
t\right)  $ have been filled out.

In the propagator~(\ref{eq:PropWignerGeneral}) for harmonic interactions of
the form~(\ref{eq:Vquad}), only terms linear in $\xi_{a,b}$ survive in the
exponent, and one is left with%
\begin{multline}
\underset{\text{quad}}{K_{w}}\left(  x_{b},p_{b},t_{b}|x_{a},p_{a}%
,t_{a}\right)  =\frac{mW}{\left(  2\pi\hbar\right)  ^{2}h\left(  t_{b}%
,t_{a}\right)  }\times\\
\times\int\exp\left(  \frac{i}{\hbar}\xi_{b}\left(  -p_{b}+\frac{m}{h\left(
t_{a},t_{b}\right)  }\left(  x_{a}W+x_{b}\frac{dh\left(  t_{a},t_{b}\right)
}{dt_{b}}-\int_{t_{a}}^{t_{b}}\frac{b\left(  s\right)  }{m}h\left(
t_{a},s\right)  \,\mathrm{d}s\right)  \right)  \right)  \,\mathrm{d}\xi_{b}\\
\times\int\exp\left(  \frac{i}{\hbar}\xi_{a}\left(  p_{a}-\frac{m}{h\left(
t_{a},t_{b}\right)  }\left(  x_{a}\frac{dh\left(  t_{a},t_{b}\right)  }%
{dt_{a}}-x_{b}W+\int_{t_{a}}^{t_{b}}\frac{b\left(  s\right)  }{m}h\left(
s,t_{b}\right)  \,\mathrm{d}s\right)  \right)  \right)  \,\mathrm{d}\xi_{a},
\end{multline}
where the remaining integrals are $\delta$ functions. Taking the results for
the boundary velocities into account, one thus readily finds%
\begin{equation}
\underset{\text{quad}}{K_{w}}\left(  x_{b},p_{b},t_{b}|x_{a},p_{a}%
,t_{a}\right)  =\delta\left(  x_{b}-x_{\text{cl}}\left(  t_{b}|x_{a}%
,p_{a},t_{a}\right)  \right)  \delta\left(  p_{b}-p_{\text{cl}}\left(
t_{b}|x_{a},p_{a},t_{a}\right)  \right)  ),
\end{equation}
which is the desired result~(\ref{eq:Harmonicexample}).

%

\end{document}